\documentclass[epsf]{mn2e}

\usepackage{epsf}
\usepackage{harvard}

\def\fd{\hbox{$.\!\!^{\rm d}$}}

\title[High Speed Photometry of SDSS J013701.06-091234.9]{High Speed Photometry of SDSS J013701.06-091234.9}
\author[M.L. Pretorius, et al.]{M.L. Pretorius$^1$\thanks{E-mail: retha@mensa.ast.uct.ac.za}, P.A. Woudt$^1$\thanks{E-mail: pwoudt@circinus.ast.uct.ac.za}, B. Warner$^{1}$\thanks{E-mail: warner@physci.uct.ac.za}, G. Bolt$^2$\thanks{E-mail: gbolt@iinet.net.au}, J. Patterson$^3$\thanks{E-mail: jop@astro.columbia.edu}, E. Armstrong$^3$\thanks{E-mail: armstrong@astro.columbia.edu} \\ 
$^1$Department of Astronomy, University of Cape Town, Private Bag, Rondebosch 7700, South Africa\\
$^2$Center for Backyard Astrophysics (Perth), Camberwarra Drive, Craigie, Western Australia\\
$^3$Department of Astronomy, Columbia University, 538 West 120th Street, New York, New York 10027}

\begin{document}



\maketitle

\begin{abstract}
We present high speed photometry of the Sloan Digital Sky Survey cataclysmic variable SDSS J013701.06-091234.9 in quiescence and during its 2003 December superoutburst.  The orbital modulation at $79.71\pm0.01$~min is double humped; the superhump period is $81.702\pm0.007$~min.  Towards the end of the outburst late superhumps with a period of $81.29\pm0.01$~min were observed.  We argue that this is a system of very low mass transfer rate, and that it probably has a long outburst interval.  
\end{abstract}

\begin{keywords}
techniques: photometric -- binaries:close -- stars: dwarf novae -- stars: individual: SDSS J013701.06-091234.9, cataclysmic variables.
\end{keywords}

\section{Introduction}
Cataclysmic variable stars (CVs) are identified in the Sloan Digital Sky Survey by spectroscopic classification of candidates selected on the basis of their colours, see \citeasnoun{sdss1}.
The discovery of SDSS J013701.06-091234.9 (hereafter abbreviated SDSS0137) is announced in the second CV release of the Sloan Digital Sky Survey \cite{sdss2}.  The optical spectrum shows a blue continuum with broad absorption around narrow emission cores of H$\beta$ and higher members of the Balmer series.  There is also some indication of broad TiO absorption features characteristic of an M dwarf primary \cite{sdss2}.  These features indicate that it is a very low mass transfer rate ($\dot{M}$) system where the accretion disc is so faint that most of the system's luminosity comes from the stars themselves.  

\citeasnoun{sdss2} obtained two hours of time resolved spectroscopy and reported an orbital period ($P_{orb}$) of 84~min; the short $P_{orb}$ is also consistent with a low $\dot{M}$.  The presence of TiO absorption, together with our photometry, indicates that this spectrum was taken during the normal quiescent state, rather than, say, during a decline from outburst, where the hydrogen absorption may originate from the disc.

We observed SDSS0137 at a quiescent $V$ magnitude near 18.6 on seven nights, the system then brightened by $\sim$6~mag and strong superhumps were seen.  This qualifies SDSS0137 as an SU UMa type dwarf nova, see e.g.  \citeasnoun{suuma} for a review of SU UMa stars.

In Section \ref{sec:obs} we present our observations in quiescence and outburst.  These results are discussed in Section \ref{sec:discussion} and summarized in Section \ref{sec:summ}.

\section{Observations}
\label{sec:obs}
We obtained photometry of SDSS0137 at the Sutherland site of the South African Astronomical Observatory (SAAO) using the 1.0- and 1.9-m telescopes together with the University of Cape Town charge coupled device (UCT CCD), \citeasnoun{uctccd} gives a description of the UCT CCD photometer and its reduction software.  In addition, the SAAO CCD was used on three nights; unlike the UCT CCD, this is not a frame transfer CCD and hence there are a few seconds of deadtime between integrations.  

All the quiescent observations were made in white light; with the UCT CCD this gives photometry with an effective wavelength similar to Johnson $V$, but with a very broad bandpass.  This, together with the nonstandard flux distribution of a CV, means that it is not possible to transform the observations onto any standard photometric system.  By observing hot subdwarf and white dwarf standards, we are able to give $V$ magnitudes that are accurate to only $\sim$0.1~mag.

Most of the outburst observations were made with telescopes of the Center for Backyard Astrophysics (CBA), see \citeasnoun{cba}.

\subsection{Quiescent light curves}
\begin{figure}
\begin{center}
\epsfxsize=8.3cm
\epsffile{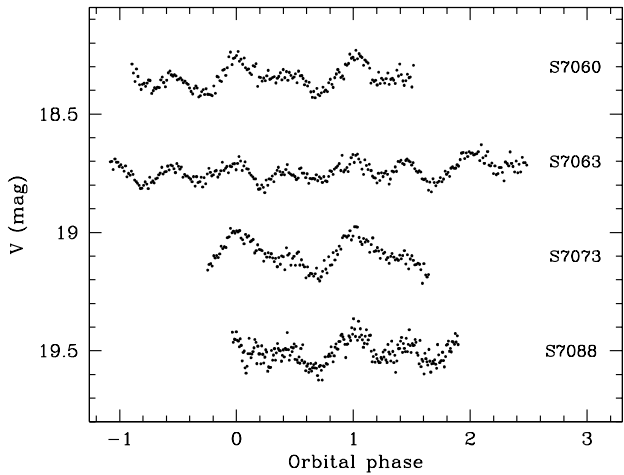}
\caption{The first four of the light curves of SDSS0137 in quiescence, phased using equation \ref{eq:eph}.  Runs S7063, S7073, and S7088 are shifted along the horizontal axis by 17, 53, and 169 cycles and down by 0.35, 0.60, and 1.00~mag respectively for display purposes.}
\label{fig:fig1}
\end{center}
\end{figure}

SDSS0137 was observed in quiescence on seven nights spread over almost four months, Table \ref{tab:qlog} is a log of these observations.  We find a double humped orbital modulation, but no coherent signals on shorter time scales.  The light curves are shown in Fig.~\ref{fig:fig1}.  Note that the relative strength of the fundamental and first harmonic changes from night to night, in S7073, e.g., the first harmonic is only barely present.  

The lower scatter on the first three nights is caused in part by the fact that the SAAO CCD (used on these nights) gives a much larger field of view, and hence includes a larger number of comparison stars, than the UCT CCD, resulting in higher quality differential photometry.  However, at least some of the scatter in run S7088 is caused by intrinsic flickering.  The UCT CCD's quantum efficiency peaks at $\lambda\sim600$~nm, but is still 25\% at $\lambda\sim350$~nm \cite{uctccd}, whereas the SAAO CCD has very poor blue and ultraviolet (but high red and near infrared) sensitivity; see figure 17 of \citeasnoun{surv2}.  Therefore the UCT CCD observations are more receptive of short wavelength flickering.

\begin{figure}
\begin{center}
\epsfxsize=8.3cm
\epsffile{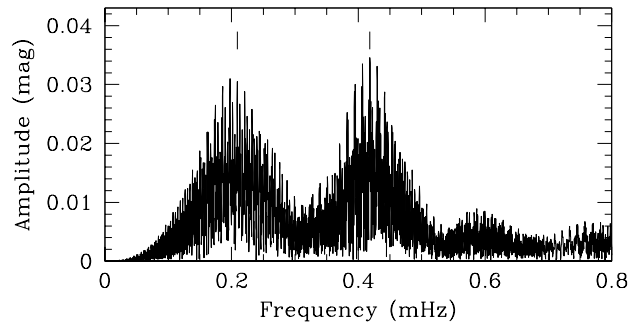}
\caption{The Fourier transform of runs S7060, S7063, S7073 and S7088 combined.  There is only one choice of alias that gives the correct fundamental and first harmonic relation, these peaks are marked by vertical bars at 0.2091~mHz (79.71~min) and 0.4183~mHz (39.85~min).}
\label{fig:fig2}
\end{center}
\end{figure}

Fig. \ref{fig:fig2} is a Fourier transform of the first four runs, the fundamental and first harmonic give an unambiguous $P_{orb}=79.71\pm0.01$~min.  The ephemeris for the times of maximum is 
\begin{equation}
HJD_{max}=2452879.594+0\fd055351(\pm7)E.
\label{eq:eph}
\end{equation}

\begin{figure}
\begin{center}
\epsfxsize=8.3cm
\epsffile{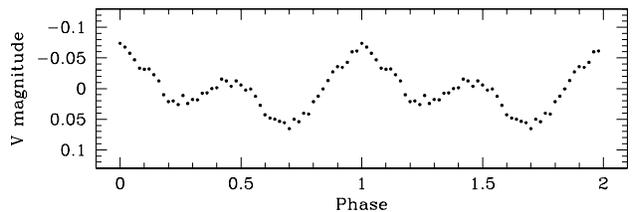}
\caption{The mean quiescent light curve of SDSS0137 (plotted twice) using the first four runs folded on the ephemeris given in equation 1.}
\label{fig:fig3}
\end{center}
\end{figure}

Fig. \ref{fig:fig3} shows the average light curve obtained by folding runs S7060, S7063, S7073 and S7088 on this ephemeris after removing a linear trend from each of the light curves individually.

\begin{table*}
 \centering
 \begin{minipage}{140mm}
  \caption{Log of the observations in quiescence.}
  \begin{tabular}{@{}lp{2.4cm}p{2.5cm}llll@{}}
  \hline
  Run No. & Date (at the start of the night) & HJD of first observation (2452000\,+) & Length (h) & $t_{int}$ (s) & Telescope & V (mag)   \\
 \hline
 S7060 & 2003 Aug. 27 & 879.54482 & 3.20 & 40 & 1.0-m & 18.4$^*$ \\
 S7063 & 2003 Aug. 28 & 880.47545 & 4.74 & 40 & 1.0-m & 18.4$^*$ \\
 S7073 & 2003 Aug. 30 & 882.51423 & 2.51 & 40 & 1.0-m & 18.5$^*$ \\
 S7088 & 2003 Sept. 5 & 888.55917 & 2.57 & 40 & 1.0-m & 18.5     \\
 S7144 & 2003 Oct. 5  & 918.44016 & 1.74 & 30 & 1.9-m & 18.6     \\
 S7168 & 2003 Dec. 16 & 990.28114 & 2.32 & 60 & 1.0-m & 18.5     \\
 S7172 & 2003 Dec. 17 & 991.27868 & 2.27 & 60 & 1.0-m & 18.6:    \\
\hline
\label{tab:qlog}
\end{tabular}
{\footnotesize
\newline
Notes: $t_{int}$ is the integration time; `:' denotes an uncertain value; $^*$taken with the SAAO CCD. \hfill}
\end{minipage}
\end{table*}

\subsection{Outburst observations}
In a system with $\dot{M}$ as low as indicated by the the spectroscopic appearance of SDSS0137, one would expect to see only fairly infrequent dwarf nova outbursts.  Nevertheless, after leaving SDSS0137 at $V=18.6$~mag on Dec. 17, we were lucky enough to find it at $V=12.5$~mag four nights later.  Table \ref{tab:oblog} is a log of the observations during the outburst.  This is the first recorded outburst.

Smooth superhumps (with a peak to peak range of 0.27~mag) were present on Dec. 21 already, implying that this was a superoutburst, that therefore SDSS0137 is an SU UMa type dwarf nova, and that the start of outburst was probably two or three days earlier.

\begin{table}
 \centering
{\small
 \begin{minipage}{80mm}
  \caption{Log of the outburst observations.}
  \begin{tabular}{@{}p{1.6cm}llll@{}}
  \hline
  HJD of first observation (2452000\,+) & Length (h) & $t_{int}$ (s) & Telescope & V (mag)   \\
 \hline
  995.29032 & 2.10 & 5  & 1.00-m  &        \\
  995.57788 & 4.93 & 2  & 1.30-m  & 12.4   \\
  996.02674 & 3.48 & 15 & 0.25-m  &        \\
  996.27143 & 0.96 & 6  & 1.00-m  &        \\
  996.63321 & 2.70 & 5  & 1.30-m  & 12.6   \\
  997.01334 & 3.97 & 15 & 0.25-m  &        \\
  997.29209 & 3.76 & 28 & 0.36-m  &        \\
  997.46431 & 3.62 & 80 & 0.25-m  & 12.7   \\
  998.00842 & 4.16 & 15 & 0.25-m  & 12.8   \\
  999.01274 & 3.81 & 30 & 0.25-m  & 12.8   \\
 1000.08011 & 2.21 & 30 & 0.25-m  &        \\
 1000.61812 & 1.75 & 45 & 0.36-m  & 13.0:  \\
 1001.45554 & 3.90 & 80 & 0.25-m  &        \\
 1001.56973 & 2.33 & 60 & 0.36-m  & 13.0:  \\
 1002.29981 & 3.21 & 28 & 0.36-m  &        \\
 1002.45077 & 3.57 & 80 & 0.25-m  &        \\
 1002.56369 & 4.24 & 5  & 1.30-m  & 13.1   \\
 1003.27273 & 1.93 & 6  & 1.00-m  & 13.1   \\
 1004.25491 & 2.60 & 28 & 0.36-m  & 13.2   \\
 1006.23953 & 0.88 & 60 & 0.41-m  &        \\
 1007.08293 & 1.63 & 30 & 0.25-m  & 13.5   \\
 1008.01126 & 3.03 & 30 & 0.25-m  & 13.6   \\
 1011.01927 & 3.13 & 30 & 0.25-m  &        \\
 1011.30694 & 0.87 & 28 & 0.36-m  & 13.9   \\
 1012.01266 & 2.87 & 60 & 0.25-m  & 14.7   \\
\hline
\label{tab:oblog}
\end{tabular}
{\footnotesize
\newline
Notes: $t_{int}$ is the integration time; the magnitudes are nightly averages; `:' denotes an uncertain value; most observations were in white light. \hfill}
\end{minipage}
}
\end{table}

\begin{figure*}
\begin{center}
\epsfxsize=16cm
\epsffile{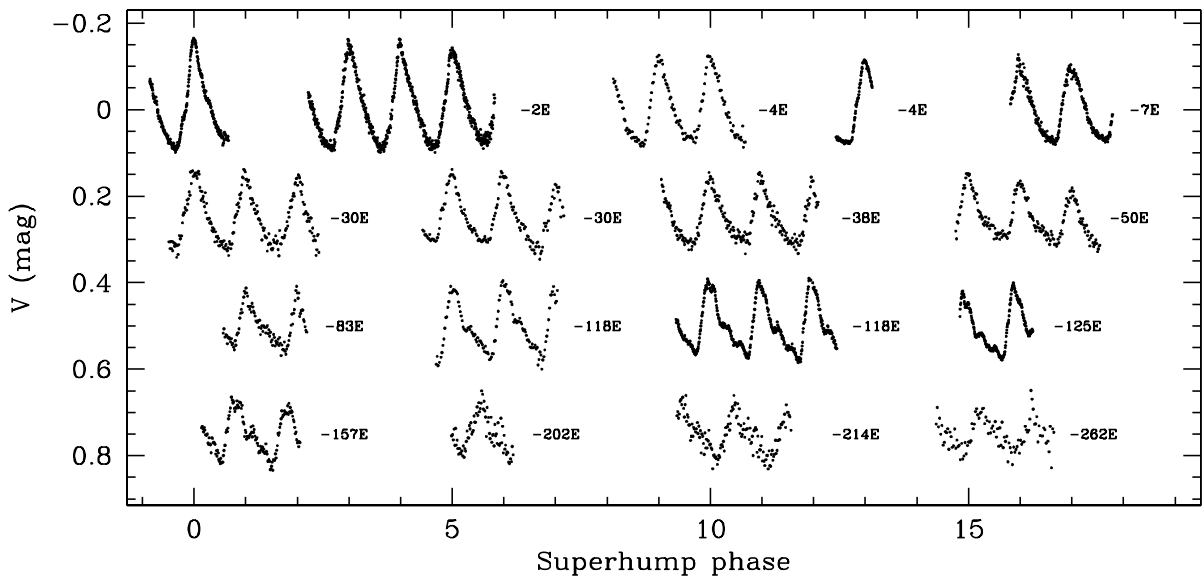}
\caption{The outburst light curves after linear trends were removed from some of them; most of these light curves were binned to improve signal to noise.  The observations are phased according to the superhump period of 81.702~min.  Individual runs are shifted along the horizontal axis by the number of cycles indicated, and arbitrarily displaced vertically.  A few of the lower quality runs are not shown.}
\label{fig:fig4}
\end{center}
\end{figure*}

\begin{figure}
\begin{center}
\epsfxsize=8.3cm
\epsffile{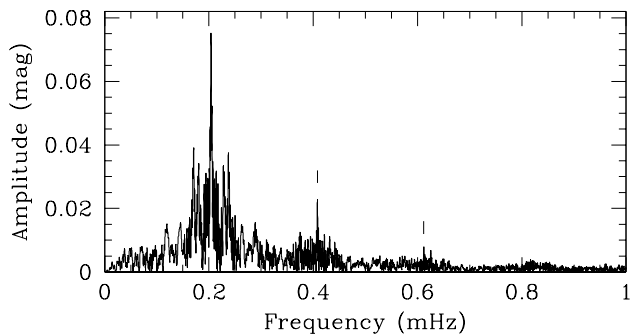}
\caption{The Fourier transform of the first eight nights of the outburst.  Vertical bars at 40.854~min and 27.231~min mark the first and second harmonic.}
\label{fig:fig5}
\end{center}
\end{figure}

\begin{figure}
\begin{center}
\epsfxsize=8.3cm
\epsffile{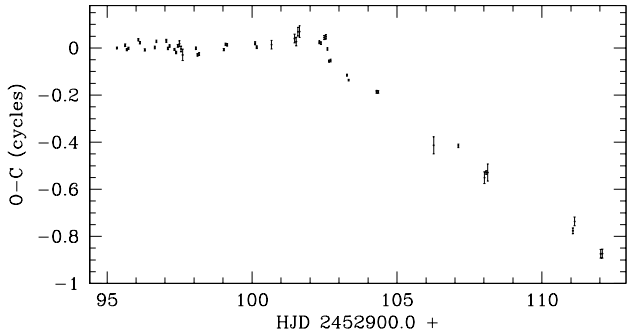}
\caption{O--C diagram computed by comparing the observed times of maximum with the times calculated from a sine function with a period of 81.702~min.  This is consistent with a constant period up until HJD 2453003, after that the phase starts changing.}
\label{fig:fig6}
\end{center}
\end{figure}

\begin{figure}
\begin{center}
\epsfxsize=8.3cm
\epsffile{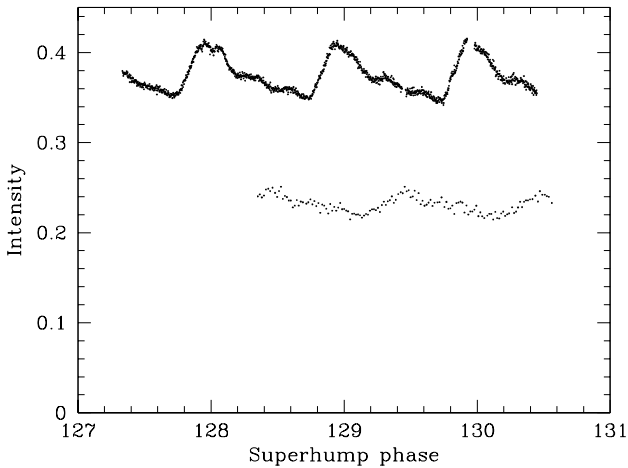}
\caption{Superhump and late superhump profiles plotted in intensity units.  There is no vertical displacement between the two light curves, they are phased using the superhump period and the time of the first observed superhump maximum.  The bottom light curve is transposed along the horizontal axis by 95 superhump cycles.}
\label{fig:fig7}
\end{center}
\end{figure}

The outburst light curves are shown in Fig. \ref{fig:fig4} (a few noisy runs are omitted).  Linear trends were removed from most of these light curves.  Combining all the data from the first eight nights of the outburst gives the Fourier transform displayed in Fig. \ref{fig:fig5}.  The superhump period ($P_{sh}$) is 81.702$\pm$0.007~min; the first and second harmonic are also clearly detected.  The light curves in Fig. \ref{fig:fig4} are phased using this period.

By the ninth night the superhumps were becoming noticeably phase shifted, another four nights later they had shifted by $\sim$0.5 cycle and could be classified as late superhumps \citeaffixed{lsh}{e.g.}.  The observed minus calculated (O--C) diagram (Fig. \ref{fig:fig6}) shows the transition from superhumps to late superhumps; there is no cycle count ambiguity in this diagram.  

Fig. \ref{fig:fig7} shows the observations on HJD\,=\,2453008 and one of the HJD\,=\,2453002 light curves plotted on the same intensity scale.  The amplitude and position of the late superhump maximum indicates that it did not develop out of anything visible in the superhump profile.  Rather, the structure in the superhump is its own repetitive profile.  

\begin{figure}
\begin{center}
\epsfxsize=8.3cm
\epsffile{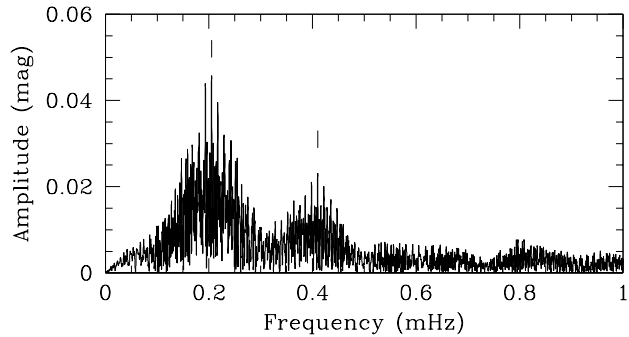}
\caption{The Fourier transform of all observations of the late superhump.  Peaks corresponding to the fundamental and first harmonic are marked.}
\label{fig:fig8}
\end{center}
\end{figure}

Fig. \ref{fig:fig8} is a Fourier transform of all the data taken after HJD\,=\,2453004; it gives a late superhump period of $81.29\pm0.01$~min.

All the outburst light curves were searched for rapid quasi-coherent oscillations, but with no success.

\section{Discussion}
\label{sec:discussion}
It has often been noted that the standard model of CV evolution predicts the existence of a large number of short period, low $\dot{M}$ CVs, but that only very few such systems are known \citeaffixed{kolb}{e.g.}.  Low $\dot{M}$ CVs are intrinsically faint, but many are being discovered in surveys that reach faint limits, such as the Sloan Digital Sky Survey.

The orbital modulation of SDSS0137 is very similar to the double hump in WZ Sge \citeaffixed{wzsge1}{e.g.}.  Three more examples of CVs with double humped orbital modulations are WX Cet \cite{wxcet}, SDSS~J161033.64-010223.3 \cite{sdss1610}, and HS 2331+3905 \cite{a-b}.  WZ~Sge and WX Cet are both low $\dot{M}$ dwarf novae of long outburst interval; no outburst has yet been observed in SDSS J161033.64-010223.3.

The double humped modulation in WZ Sge (and the other systems like it) has not been convincingly modeled, but is probably caused by the bright spot.  An early explanation was that the bright spot has a large scale height and is thus visible above the inner accretion disc when it is on the far side of the disc (\citename*{smak} \citeyear*{smak}; \citename*{wzsge2} \citeyear*{wzsge2}).  Since then, it has been demonstrated that the disc in WZ Sge is optically thin (\citename*{skid1} \citeyear*{skid1}; \citename*{skid2} \citeyear*{skid2}) so that the bright spot can be seen through it. 

The second hump in SDSS0137 has lower amplitude during run S7073 than in any of our other runs.  This has also been observed in WZ Sge \cite{smak}; in WX Cet the second hump is sometimes not detected at all \cite{wxcet}.  A possible explanation for this is that the optical thickness of the disc can temporarily increase so that the spot is (at least partially) obscured when it is on the side of the disc facing away from the observer \cite{smak}.  However, in the case of our run S7073, an increase in optical depth of the disc can't have been caused by an increase in $\dot{M}$ through the disc, since the system was 0.10~mag fainter than in run S7063 and 0.14~mag fainter than in run S7060.
Although there are a few features in the spectrum which may be attributed to the secondary, it does not contribute a large fraction of the total flux, therefore it seems very unlikely that the orbital modulation contains a substantial amount of ellipsoidal variation from the secondary.

Normal superhumps are thought to be caused by tidal stresses acting on an eccentric disc \citeaffixed{wh}{e.g.}.  Repetitive structure in superhump profiles similar to what can be seen in Fig. \ref{fig:fig7}, and in more of the light curves in Fig. \ref{fig:fig4}, has been observed in VW Hyi as well (\citename*{sv} \citeyear*{sv}; \citename*{brian85} \citeyear*{brian85}).  The superhump's fractional period excess is $\epsilon=(P_{sh}-P_{orb})/P_{orb}=0.0250\pm0.0001$.

The interval between supermaximum and the first appearance of superhumps is less than 4 days.  This indicates that the outburst interval in SDSS0137 is probably years at most, rather than decades \cite{bible}.

A variation in bright spot brightness, resulting from the varying depth in the white dwarf potential well at which the stream impacts the disc, was initially proposed as a model for superhumps \cite{vogt82}, but has since been used to explain late superhumps (e.g. \citename*{wh} \citeyear*{wh}; \citename*{murr} \citeyear*{murr}; \citename*{rolfe} \citeyear*{rolfe}), but see also \citeasnoun{hess}.  \citeasnoun{hellier} proposes an alteration to the tidal thermal instability \cite{tti}, namely that the tidal and thermal instabilities are uncoupled.  The tidal eccentricity can persist after the disc has come down from the thermal high state at the end of the outburst.  The viscous dissipation -- caused by tidal stresses -- which results in superhumps, then decreases because the disc viscosity is now much lower.  At this point the late superhumps should become detectable above the superoutburst light.  It is therefore perhaps surprising that the late superhump can not be seen in the HJD\,=\,2453002 light curves.    

There are two other systems for which the transition from superhumps to late superhumps has been carefully tracked (V1159 Ori \cite{v1159} and IY UMa \cite{iyuma}).  In both these examples O--C increases during the transition, in contrast to what is shown in Fig. \ref{fig:fig6}.

\section{Summary}
\label{sec:summ}
We classify the CV SDSS0137 as an SU UMa type dwarf nova and report an orbital period of 79.71~min and a superhump period of 81.702~min.  SDSS0137 has a double humped orbital modulation, which is compatible with the very low $\dot{M}$ indicated by the spectral appearance.  Late superhumps developed $\sim$11 days after the start of the outburst.

\section*{Acknowledgments}
BW's research is funded by the University of Cape Town.  PAW is supported by the National Research Foundation and by funds made available to BW from the University of Cape Town. MLP is funded by the Department of Labour.  We are very grateful to Berto Monard, David Messier, Don Starkey and Arto Oksanen for contributing observations.  The paper has benefited from constructive comments by the referee, Dr. Boris G\"{a}nsicke.

\end{document}